\documentstyle[12pt]{article}

\begin{document}
\begin{center}
{\Large \bf On the Abnormal Type
Anomalous Solutions of Quasipotential Equations}
\medskip

{\bf Anzor A.\ Khelashvili, George A.\ Khelashvili,
Nicholoz Kiknadze and 
Temur P.\ Nadareishvili\footnote{Electronic address:temo@hepi.edu.ge}}
\medskip

{\it High Energy Physics Institute and Tbilisi State University, Tbilisi, Georgia}
\end{center}

\begin{abstract}
It is shown that there exist solutions of the quasipotential equation
exhibiting the abnormal type behaviour of the Bethe-Salpeter
equation.
\end{abstract}
\medskip

It is well known that the relativistic Bethe-Salpeter (BS) equation has not only ordinary (so called normal), but 
also the abnormal solutions \cite{1} which have no non-relativistic analogue. Such solutions were discovered long 
ago in the simplest Wick-Cutkosky model \cite{2},\cite{3}. It should be noted that in spite of considerable effort 
in this direction (for details see \cite{1}) nature of the abnormal solutions is still unclear. The only thing that is 
clear is that they are related to the excitations of the relative time (or, in the momentum space, relative energy) 
degrees of freedom. Also it is established that the norm of such solutions can become negative --- the sign of the norm is 
proportional to the parity of BS amplitude with respect to the relative energy which can be both positive or 
negative. In general case it makes quantum-mechanical (probabilistic) interpretation of the BS amplitude 
problematic.

The so-called quasipotential wave-function \cite{4} is free of negative norm problem. 
The wave-function is defined as the equal-time BS amplitude. Amplitudes which are odd with respect to the relative 
energy can not survive in the equal-time limit and so 
only those with the positive norm survive. However, as in the BS equation there are the abnormal
solutions with the positive norm too and so they have to show up in the quasipotential approach as well.

An important feature of the quasipotential approach is the dependence
of the quasipotential on the total energy of
the system even if the kernel of the BS equation did not depend on it.
It's worth noting that excitations of the
relative time degree of freedom are transformed into the energy-dependence of the quasipotential, so one can  
expect existence of some ``extra" solutions. 

The above consideration is not  new \cite{4} but theoreticians started to show interest to it for practical purposes 
only recently \cite{5}--\cite{8}.

Below we will concentrate on a simple quasipotential and study the structure
of its spectrum. The quasipotential equation in the single-photon
exchange approximation in QED for wave function projected onto
the positive-frequency states lookes like \cite{5},\cite{9}:
\begin{equation}
 w_p(2w_p-M)\Psi(\vec p)=\frac{1}{(2\pi)^3}\int\frac{d\vec p}{2 w_{\vec q}}V(M;\vec p, \vec q) \Psi(\vec q) \ ,    
\label{1}
\end{equation}
where $ w_p=\sqrt{\vec p^2+m^2}$. We have used the center of mass frame and set the fermion masses to the 
same value $m$. Quasipotential $V$ depends on the total mass of system $M$ and in the $O(e^2)$ approximation 
has the form:
\begin{equation}
V(M;\vec p, \vec q)=\frac{(2me)^2}{|\vec p-\vec q|(M- w_p- w_q-|\vec p-\vec q|+io}\ .
\label{2}
\end{equation}
It describes the interaction of the fermions with the opposite charges. For the fermions
with  the same charges  we have to make the
change: $e^2\to -e^2$. On the mass-shell $M=2 w_p=2 w_q$ quasipotential (\ref{2}) coincides with the Coulomb 
potential.The quasipotential  (\ref{2}) is nonlocal function in general 
and solution of corresponding equation(\ref{1}) is very hard. In the papers \cite{6}, \cite{10} the limit of small momenta $|\vec p|, |\vec q| \ll m$ is considered. In 
this approximation the quasipotential takes the following local form:
\begin{equation}
V(M;\vec p, \vec q)= V(M; |\vec p-\vec q|)=\frac{(2me)^2}{|\vec p-\vec q|({\cal E}-|\vec p-\vec q|+io)}\ .
\label{3}
\end{equation}
Here ${\cal E}=M-2m$ is the binding energy. It should be noted
that different quasipotentials of a number of
other models reduce to the same form \cite{3} in the
small momenta approximation where, unlike the full
nonrelativistic approximation, the binding energy
is keeped (see e.g.\ \cite{11}).

Below we will investigate the small momenta approximation of the Eq.\
(\ref{1}) with quasipotential (\ref{2}) for
negative binding energies (${\cal E}=-E<0$). The resulting Schr\"odinger equation in the momentum space is:

\begin{equation}
\left( \frac{\vec p^2}{m}+E \right)\Psi(\vec p)=\frac{\alpha}{2\pi^2}\int d\vec q\frac{\Psi(\vec q)}
{|\vec p-\vec q|(E+|\vec p-\vec q|+io)}\ .
\label{4}
\end{equation}
Here $\alpha=e^2/4\pi$ is the fine structure constant and the dependence
of the potential over the binding energy
is the relic of relativity. Eq.\ (\ref{4}) can be studied by the standard
numerical methods applicable for Fredholm
type kernels but for transparency we will analyze it in the coordinate
space\cite{11} :
\begin{equation}
\left(E+\frac{\nabla^2}{m}\right)\Psi(r)=V_E(r) \Psi(r)\ ,
\label{5}
\end{equation}
where
\begin{equation}
V_E(r)=-\frac{2\alpha}{\pi}\frac{f(Er)}{r}\ ;
\label{6}
\end{equation}
\begin{equation}
f(Er)=Ci(Er)\sin(Er)-si(Er)\cos(Er)\ .
\label{7}
\end{equation}
Here $Ci(x)$ and $si(x)$ are the integral cosine and sine, respectivly.
Despite the dependence over trigonometric
functions the function $f(x)$ is a smooth and non-negative one
(See fig.\ 1 in\cite{12}  ). It has the following asymptotics \cite{13}:
\begin{eqnarray}
f(x) & \approx &
 \frac{\pi}{2}+x\ln x+(\gamma-1)x-\frac{\pi}{4}x^2+O(x^3); \qquad x\ll 1, 
\label{8} 
\\ f(x) & \approx &
 \frac{1}{x}(1-\frac{2!}{x^2}+\frac{4!}{x^4}-\cdots); \qquad
 x\gg 1. 
\label{9} 
\end{eqnarray}
Here $\gamma=0.5772\cdots$ is the Euler constant. Hence for
${\cal E}<0$ potential (\ref{6}) behaves like \cite{11}:
\begin{equation}
\left.V_E(r)\right|_{Er\ll 1}\sim-\frac{\alpha}{r}\left[1+
\frac{2}{\pi}Er\ln(Er)+\frac{2}{\pi}(\gamma-1)Er-\frac{\pi}{4}
(Er)^2\right],
\label{10}
\end{equation}
	
\begin{equation}
\left.V_E(r)\right|_{Er\gg 1}\sim-\frac{2\alpha}{Er^2}.
\label{11}
\end{equation}

Clearly, when $E=0$ the potential is pure Coulombic:
\begin{equation}
V_{E=0}(r)=-\frac{\alpha}{r}.
\label{12}
\end{equation}

So  the behavior of the potential at the infinity differs
from that of the Coulombic one if we preserve the binding energy dependence
in the potential.
The question arises precisely in this place-whether there appear ''extra'' solutions of the Schr\"odinger equation because of presence of parameter (energy) in the potential.
There are quite definite results in Quantum Mechanics for local potentials with
$r^{-2}$ asymptotics\ \cite{14}. Energy-dependent potentials were also 
investigated, but for proof usually
the limit $E=0$ is considered (see e.g.\ \cite{15}).  We can
not take that limit as the asymptotic regime of potential changes drastically, but it is not difficult to
show that main conclusions of the theorem from \cite{14} remain valid in
our case too. Indeed, let us study the
asymtotical  solution of the radial Schr\"odinger equation for our case:
\begin{equation}
\chi(r)''-\left[\kappa^2-\frac{\gamma(E)}{r^2}\right]\chi(r)=0\ , \qquad\qquad Er\gg 1.
\label{13}
\end{equation}
Here
\begin{equation}
\kappa^2\equiv mE, \qquad \gamma(E)\equiv2\alpha m/\pi E-l(l+1)
\label{14}
\end{equation}

General solution of this equation is the superposition of the Besel functions with imaginary arguments:
\begin{equation}
\chi(r)=c_1\sqrt{r}J_{\sqrt{\frac{1}{4}-\gamma(E)}}(i\kappa r)+ c_2\sqrt{r}J_{-\sqrt{\frac{1}{4}-
\gamma(E)}}(i\kappa r), \qquad r\gg 1/E\ .
\label{15}
\end{equation}

As long as we are interested in finding bound states we must choose the coefficients $c_{1,2}$ so that to get 
vanishing behavior at spatial infinity. Such choice corresponds to the McDonald
function $K_\nu$ with
\begin{equation}
c_1=-c_2=-\frac{\pi}{2sin(\pi\nu)},\qquad\qquad \nu\equiv \sqrt{\frac{1}{4}-\gamma(E)}
\label{16}
\end{equation}
and
\begin{equation}
\chi(r)\approx\sqrt(r)K_{\sqrt{\frac{1}{4}-\gamma(E)}}(\kappa r)\ , \qquad\qquad r\gg \frac{1}{E}
\label{17}
\end{equation}

It is well known that the McDonald function has zeroes on real axis only
for imaginary indices. So there are the
following alternatives:

\begin{description}
\item{(a)} if $\gamma(E)<1/4$ the radial function $\chi(r)$ has no zeroes at the infinity, i.e.\ the Schr\"odinger 
equation with the potential (\ref{6}) has none or finite number of discrete levels;
\item{(b)} if $\gamma(E)>1/4$ the index of the function (\ref{17}) is
imaginary and the radial function has
infinite number of  zeroes at the spatial infinity, i.e.\ there is infinite
number of discrete levels.
\end{description}

Let us now discuss physical consequences. According to {\bf (b)} the discrete levels may arise
only (see \cite{16}) for
\begin{equation}
\frac{2\alpha m}{\pi E}>\frac{1}{4}+l(l+1)\qquad \mbox{or} \qquad \alpha>\frac{\pi E}{8m}+l(l+1) \frac{\pi 
E}{2m}\ .
\label{18}
\end{equation}

At the first sight this condition can be satisfied for usual values of $ \alpha, \alpha \approx 1/137$.
E.g.\ in the zone of the
Balmer series where $E\approx m\alpha^2$ the left-hand-side of the
first inequality in (\ref{18}) is of the order of $\alpha^{-1}$ and so this
inequality can be satisfied.
However, careful examination of the location of zeroes of the function $K_{\nu}(x)$ reveals opposite situation. 
Indeed, according to the theorem {\it 7.13.9} \cite{16} the smallest real positive zero of the function 
$K_{\nu}(x)$ for imaginary $\nu$ lies in the region $x<|\nu|$.
All asymptotic formulae \cite{16} lead to
the same conclusion for any other zeroes. So,
together with $\gamma(E)>1/4$ we have to impose also the
following condition:
$$
\kappa r<\sqrt{\gamma(E)-\frac{1}{4}}
$$
or, equvalently
\begin{equation}
\alpha>\frac{\pi E}{2m}[\frac{1}{4}+l(l+1)]+\frac{\pi}{2}(Er)^2 \ .
\label{19}
\end{equation}
But as long as $Er\gg 1$ this new condition can not be satisfied for small reasonable 
values of $\alpha$. 

Thus, study of the asymptotic behavior of the Schr\"odinger equation
indicates that existence of any 
discrete levels apart from the Coulombic ones is possible only for large values of the coupling $\alpha$,i.e.\ in the region where the single-photon 
approximations becomes invalid.It is obvios,therefore,that discrete levels were not found in numerical solutions of this equation in \cite{5}.  It should be noted that the described picture is quite similar to the anomalous 
solutions of the BS equation. It is essential that when
we neglect the binding energy in quasipotential, the regime of asymptotic behavior of the
potential changes discountinously and the anomalous solutions disappear. 
Therefore their appearence is the consequence of the off-energy-shell effects.  We are inclined to think that the main
features of the results described above are valid
also for non-local quasipotential (\ref{2}).
So we have showed that the above mentioned quasipotential equation is not free from
the pathologies of the BS equation such as the Abnormal solutions.

\end{document}